\documentclass[journal,10pt]{IEEEtran}
\usepackage{cite}      
\usepackage{graphicx}  
\usepackage{url}
\usepackage{times}
\usepackage{amsmath}   
\usepackage{pdflscape}
\usepackage{amsfonts,amsthm,amssymb}
\usepackage[normalem]{ulem}
\usepackage{mathtools}
\usepackage{nicefrac}
\usepackage[none]{hyphenat}
\usepackage{array}
\usepackage{fontenc}
\usepackage{color}
\usepackage{etoolbox}
\usepackage{bm}
\usepackage{comment}
\usepackage{caption}
\usepackage{subcaption}
\usepackage{amsmath,amsthm,amssymb, amsfonts}
\usepackage{amssymb,mathtools}

\usepackage {algorithm, algorithmic}
\usepackage{nomencl}
\renewcommand{\baselinestretch}{0.87}

\usepackage{accents}

\newcommand{\argmax}{\arg\!\max}

\newlength{\dhatheight}

\DeclareMathOperator{\xt}{\mathbf{{x}}}
\DeclareMathOperator{\xr}{\mathbf{\hat{\hat{x}}}}
\DeclareMathOperator{\xc}{\mathbf{\hat{x}}}
\DeclareMathOperator{\kd}{\vardbtilde{\mathcal{K}}}

\newcommand{\vardbtilde}[1]{\tilde{\raisebox{0pt}[0.85\height]{$\tilde{#1}$}}}
\usepackage{xparse}

\begin{document}

\setlength{\textfloatsep}{0pt}
\setlength{\intextsep}{10pt plus 2pt minus 2pt}

\title{\Large Radar Enhanced Multi-Armed Bandit for Rapid Beam Selection in Millimeter Wave Communications \vspace{-2mm}}

\author{
	\IEEEauthorblockN{Akanksha~Sneh\textsuperscript{1}, Sumit~Darak\textsuperscript{1}, Shobha~Sundar~Ram\textsuperscript{1} and Manjesh Hanawal\textsuperscript{2} \\
\textsuperscript{1}Indraprastha Institute of Information Technology Delhi, New Delhi 110020 India\\
\textsuperscript{2}Indian Institute of Technology Bombay, Mumbai 400076 India\\
E-mail: \{akankshas, sumit, shobha \}@iiitd.ac.in}; 
mhanawal@iitb.ac.in}

\maketitle

\begin{abstract}
Multi-arm bandit (MAB) algorithms have been used to learn optimal beams for millimeter wave communication systems. Here, the  complexity of learning the optimal beam linearly scales with the number of beams, leading to high latency when there are a large number of beams. In this work, we propose to integrate radar with communication to enhance the MAB learning performance by searching only those beams where the radar detects a scatterer. Further, we use radar to distinguish the beams that show mobile targets from those which indicate the presence of static clutter, thereby reducing the number of beams to scan. Simulations show that our proposed radar-enhanced MAB reduces the exploration time by searching only the beams with distinct radar mobile targets resulting in improved throughput. 
\end{abstract}


\section{Introduction}
\label{sec: Intro}
Millimeter wave (mmW) unlicensed spectrum has been identified as a viable solution for realizing high data rate communications between connected vehicles \cite{USspectrum,Canadaspectrum,Europespectrum,Japanspectrum,SKspectrum}. The communication links are, however, characterized by high atmospheric absorption and hence can be operational only in short-range line-of-sight scenarios with highly directional beams realized through analog or digital beamforming at the transmitter/receiver. Digital beamforming allows for multiple simultaneous beams but is costly and complicated to implement since multiple phase and time-synchronized RF/mmW chains are required \cite{yang2018digital}. Analog beamforming is less costly since it involves a single beam at a time. But there is considerable overhead expended by the communication protocol and a long search/exploration time to scan the entire field of view  and select the best beams for each mobile user (MU). This results in high latency and shorter service/exploitation time available for communication causing low throughput. 

There have been several recent works that have applied multi-armed bandit (MAB) algorithms for reducing the exploration time of the best beams in order to increase the exploitation time for subsequent mmW communications \cite{zhang2020beam,booth2019multi,aykin2020mamba,chafaa2019adversarial}. 
MAB algorithms are a class of algorithms within the reinforcement learning framework which provides a basis for making decisions under uncertainty. MAB-based beam selection works, such as \cite{booth2019multi,aykin2020mamba}, have relied on a strategy where the base station (BS) waits for the feedback (reward) over the uplink for the beam selection in subsequent time slots. In such time-slotted communication, the transmitter can switch the beam only once in a slot, and the duration of each slot depends on the time taken by the MU to process the downlink signal and share the feedback over the uplink. 


\indent The use of radar signals for detecting the presence of targets can potentially speed-up the beam search. However, there are certain challenges in the integration of radar with communication physical layer. The use of an auxiliary radar sensor for detecting MU, cannot be considered, as it would increase the cost and complexity of the system. Further, the radar and communication functionalities would have to be synchronized as well as managed for interference. Instead, we propose that an integrated sensing and communication system be utilized for mmW communication such as those proposed by \cite{grossi2018opportunistic,kumari_ieee_2018}. Here, a common waveform on a common spectrum is used for joint radar sensing and communications. Hence, no separate hardware/spectrum/synchronization or interference management is required to support both functionalities.  Note that joint radar communication (JRC) systems have been explored over the last several decades to tackle spectral congestion issues \cite{liu2020joint}. While some works have studied how to manage mutual interference that arises from the coexistence of both systems on a common spectrum  \cite{deng2013interference}, others have exploited the communication signal as an opportunistic illumination for passive radar receivers \cite{zeng2016wireless}. We identify our work to belong to the third category of research that explores the collaborative design of JRC systems to improve the performance of each functionality \cite{Hassanien2019dual,duggal2020doppler}.

In this work, we propose incorporating a radar sensing mechanism into the MAB framework at the BS to overcome the limitations listed above. In the proposed framework, the radar at the BS is used to detect the presence of MU in the candidate beams based on the strength of the scattered signal (\emph{amplitude gated radar enhanced MAB}) and the Doppler frequency shift (\emph{Doppler gated radar enhanced MAB}) introduced to the radar signal. Only those beams that indicate the presence of a mobile radar target will be further scanned for the presence of a MU. Radar detection-based decision-making will be much faster than communication metric-based decision-making due to multiple reasons. First, the feedback for a radar signal is nearly instantaneous since it is based on the electromagnetic scattering of the signal by mobile targets.  Second, the exploration time is substantially reduced by restricting the number of candidate beams that have to be scanned. Due to these factors, the overall exploration time will be reduced resulting in rapid beam alignment and improved overall communication throughput. \\
\indent\emph{Notation:} In our paper, scalar variables, vectors, and matrices are denoted with regular, and boldface lower and upper case characters respectively. 
Vector superscript $T$ and symbol $\otimes$ denote transpose and convolution operations. 
\section{Rad-Com Signal Model}
\label{sec:Signal Model}
We first present the signal models for the JRC transmitter and receiver based on the IEEE 802.11ad protocol where the Golay sequences in the communication frame are exploited for radar sensing  \cite{grossi2018opportunistic,kumari_ieee_2018,duggal2020doppler}. The digital waveform $\xt_q[m]$ corresponds to the Golay sequence in the $q^{th}$ packet transmitted at a pulse repetition interval of $T_P$ with $m=1,2,\ldots,M$ samples.
These digital packets are then converted into analog signals $\xt_q(t)$ at the BS as follows:
\par\noindent\small
\begin{align}
    \xt_q(t) =\sum_{m=0}^{M}x_q[mT_s]\delta\left(t-mT_s - (q-1)T_P\right),
\end{align}\normalsize
where $T_s$ is the sampling time. The signal is then amplified with energy $E_s$, convolved with a transmit shaping filter, $\mathbf{g}_T$, and then passed through analog upconversion to the mmW carrier frequency $f_c$ as
\par\noindent\small
\begin{align}
   \xt_{{q_{uc}}}(t) = \sqrt{E_s}\left(\xt_q(t) \otimes \mathbf{g}_T(t)\right)e^{+j 2\pi f_c t}.
\end{align} \normalsize
The upconverted signal is then transmitted via analog beamforming through a uniform linear array (ULA) of $P_{BS}$ elements after applying complex antenna weight vector at BS transmitter (BS-TX), $\mathbf{w}_{{BS}_\theta} \in \mathcal{C}^{P_{BS}\times 1}$ for a given angle $\theta$, resulting in \par\noindent\small
\begin{align}
    \mathbf{X}_{q_{uc}}(t) = \mathbf{w}_{{BS}_\theta}\mathbf{x_{q_{uc}}}^T(t).
\end{align}\normalsize
Here, $\mathbf{w}_{{BS}_\theta}= [1\;e^{-jk_c d_{BS}\sin \theta}\;\cdots \ e^{-jk_c d_{BS}(P_{BS}-1)\sin \theta}]$ where $k_c$ is the propagation constant and $d_{BS}$ is the uniform element spacing. \emph{It is important to note that the problem of searching for a new beam only arises for a MU that has changed its position and not a static user.}

\noindent\textbf{Radar received signal}: Along a pre-determined beam angle $\theta$, we assume that there are $B$ radar targets present in the channel including MU and other discrete clutter scatterers. Then the received signal at the $P_{BS}$-element ULA at the BS receiver (BS-RX) after being reflected from the targets, is 
\par\noindent\small
\begin{align}
\begin{split}
\label{eq:rx_radar}
    \xr_q(t) = \sum_{b=1}^B \sigma_b \mathbf{w}_{{BS}_\theta}\mathbf{u}_{\theta}\mathbf{H_{r}}^2\mathbf{u}^T_{\theta}\left[\mathbf{X}_{q_{uc}}(t-2\tau_b)\right] 
    + \mathbf{\rho}(t),
\end{split}    
\end{align} \normalsize
where $\tau_b$ denotes the time delay caused by one-way propagation and $\sigma_b$ is the strength of the reflection from each $b^{th}$ point target obtained from Frii's radar range equation. $\mathbf{H_{r}}^2$ is the ${P_{BS}\times {P_{BS}}}$ channel matrix that includes the direct path and multipath due to static clutter scatterers (SCS) present in the environment modeled in a manner described in \cite{heath2016overview} but for two-way propagation. The steering vector from the ULA corresponding to $\theta$ is given by $\mathbf{u}_{\theta}^T = [1\;e^{jk_c d_{BS}\sin \theta}\;\cdots \ e^{jk_c d_{BS}(P_{BS}-1)\sin \theta}]$ while $\mathbf{\rho}$ is the additive circular symmetric white Gaussian noise at the BS-RX.
We assume that each radar target is moving with a constant radial velocity $v_b$, such that the Doppler shift is $f_b = 2v_b/\lambda$ where $\lambda$ is the wavelength. 
After down-conversion and digitization, the received radar signal is 
\par\noindent\small
\begin{align}
\label{eq:rx_radar2}
 \xr_q[m] = \sum_{b=1}^B \sigma_b \mathbf{w}_{BS}\mathbf{u}_{\theta}\mathbf{u}^T_{\theta} \mathbf{X}_{q_{uc}}\left[m-m_b\right]e^{-j2\pi f_bqT_P} 
 +  \mathbf{\rho},
\end{align} \normalsize
where, $m_b$ is the sample index corresponding to $\tau_b$.

\noindent\textbf{Radar signal processing:} The radar received signal at BS-RX gathered over $Q$ packets is first converted to a radar rectangle of dimension $[M \times Q]$ which then goes to the radar signal processing block to obtain the range and Doppler of the corresponding target.
The range estimation output, $\mathbf{\chi_q}$, is obtained through the matched filtering for each $q^{th}$ packet, $ \mathbf{\chi_q} = \xr_q  \otimes \xt_q.$ The output is processed through the ordered-statistics constant false alarm (OS-CFAR) to estimate each $a^{th}$ peak with amplitude, $\hat{\sigma}_a$, at the range $\hat{r}_a$. This $\hat{\sigma}_a$ information is used subsequently for the \emph{amplitude gated radar-enhanced MAB algorithm} discussed in the next section. 
Next, Doppler estimation is carried out through one-dimensional multiple signal classification (MUSIC) for each $a^{th}$ peak across the $Q$ packets to estimate the corresponding $\hat{f}_a$. This $\hat{f}_a$ information is used for the \emph{Doppler gated radar-enhanced MAB algorithm} discussed in the next section. 

\noindent\textbf{Communication received signal and processing:} The one-way propagated communication signal, $\xc(t)$, is received  at the $P_{MU}$ element ULA at the MU receiver (MU-RX) as shown in 
\par\noindent\small
\begin{align}
\label{eq:rx_comm}
    \xc(t) =  \mathbf{w}_{MU_\phi}\mathbf{u}_{\phi}\mathbf{H_{c}}\mathbf{u}^T_{\theta}\left[\mathbf{X}_{q_{uc}}(t-\tau_b)\right] + \mathbf{\delta}(t).
\end{align}\normalsize
Here, $\mathbf{w}_{MU_\phi}$ represents the weights applied at the MU-RX and $\mathbf{u}_{\phi} = [1\;e^{jk_c d_{MU}\sin \phi}\;\cdots ]$ is the steering vector for the BS at $\phi$ for $d_{MU}$ antenna element spacing. $H_{c}$ is the one-way propagation channel matrix ${P_{BS}\times {P_{MU}}}$ model \cite{heath2016overview} and $\mathbf{\delta}$ is the additive circular-symmetric white Gaussian noise at MU-RX. The signal $\hat{{\mathbf{x}}}(t)$ is  received by MU and gets processed and the corresponding signal-to-noise ratio (SNR) is sent back to the BS as uplink feedback. Note that the processing time for the MU results in a greater delay for the uplink signal to return to the BS compared to the nearly instantaneous radar-scattered signal. Second, the uplink $\xc$ is distinguished from $\xr$ at the BS-RX through cross-correlation with $\xt$. Due to the nature of the Golay sequence, the peak-to-sidelobe ratio after cross-correlation for $\xr$ is very high compared to the $\xc$.
 \section{Proposed MAB Framework for JRC}
\label{sec:Algos}
\begin{figure*}[htbp]
    \centering
    \includegraphics[scale=0.53]{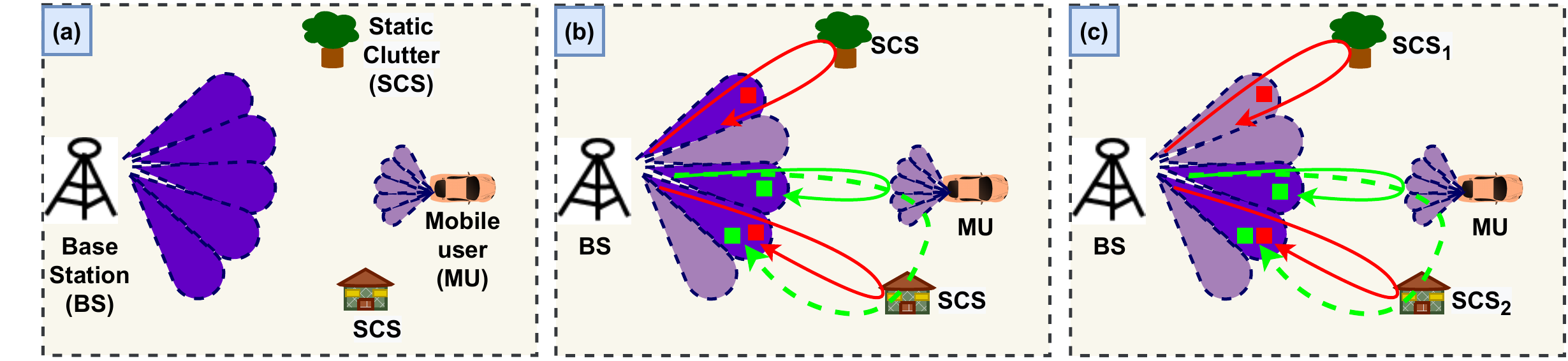}
    \vspace{-1mm}
    \caption{\footnotesize System model showing (a)  Standard non-radar based MAB beam selection, (b) Amplitude Gated Radar-Enhanced MAB, (c) Doppler Gated Radar-Enhanced MAB. Beams with red and green boxes indicate zero and non-zero Doppler targets respectively.}
    \label{fig:setup_model}
   \vspace{-4mm}
\end{figure*}

In this section, we set up the beam-selection problem between BS and MU as MAB and develop the algorithms that speed up the beam selection. The standard stochastic MAB consists of a set of $\mathcal{K}$ arms (predetermined beams) and a single player (the BS-TX/RX) as shown in Fig.\ref{fig:setup_model}a.
In each time slot, the BS-TX/RX, selects a single $k^{th}$ beam and receives the reward - the SNR of the communication link, $S_k$, obtained from uplink feedback along the beam. For each arm, the reward is assumed to be drawn independently across time from distributions that are stationary and independent across arms. 
The performance metric is equal to the difference between the SNR of the optimal beam and the SNR over the selected beam. We define this as regret which is given as \par\noindent\small
\begin{equation}
\label{Eq:regret}
    R =   T S_{k'}  - \mathop{\mathbb{E}} \Bigg[ \sum_{k\in \mathcal{K}} S_k N_k \Bigg]
\end{equation} \normalsize
where $T$ is the total number of time slots, $N_k$ is the number of times the beam $k$ is selected by BS and $k'$ =$\argmax\limits_{k \in \mathcal{K}}$ $S_{k}$. The expectation here is with respect to the random number of pulls of the arms $(N_k)$. Thus, the regret can be minimized by selecting the optimal beam, $k'$, i.e., the beam with the highest SNR as many times as possible in a given horizon of size $T$.  
In this paper, we limit our discussion to the upper confidence bound-based (UCB) MAB algorithm \cite{auer2002finite}  and provide regret bounds. The proposed idea can be easily extended to other MAB algorithms such as UCB variants and Thompson Sampling.
\vspace{-3mm}
\subsection{SNR Based Beam Selection using UCB}
\label{SS:UCB}
We first present the conventional approach for beam selection using the UCB algorithm given in Algorithm 1: $\mathbf{UCB}_{SNR}$. The communication is assumed to be time-slotted. At each $t^{th}$ time slot, the BS-TX transmits over the beam selected by the UCB algorithm. The algorithm selects each of the $\mathcal{K}$ beams once at the beginning (lines 4-6), and thereafter, the beam selection is based on the UCB index (lines 7-8). Here, $T_{RSP}$ denotes the time slots required for RSP and it is set to 0 for $\mathbf{UCB}_{SNR}$. We denote the index of the selected beam and corresponding reward, i.e., instantaneous normalized SNR, as $I_t$ and $W_t$, respectively. At the end of each time slot, the parameters are updated (line 12). The UCB index is calculated for each beam separately and it is given as \par\noindent\small
\begin{equation}
\label{equ:UCB}
    UCB_k(t) = \frac{
    \hat{S}_k}{N_k} +\sqrt{\frac{2\log(t)}{N_k}},
    \vspace{-2mm}
\end{equation}\normalsize
where $\hat{S}_k$ denotes the empirical mean of $k^{th}$ arm using samples obtained till time $t$. The expected regret of $\mathbf{UCB}_{SNR}$ scales as $\mathcal{O}(\sum_{k \in \mathcal{K}\backslash k^\prime }\frac{\log T}{\Delta_k})$ \cite{bubeck2012regret} where $\Delta_k=S_{k^\prime}-S_k$ for all $k\neq k^\prime$. Furthermore, it suffers from high exploration time especially when $\mathcal{K}$ is large. Both these drawbacks limit the usefulness of $\mathbf{UCB}_{SNR}$ for mmW communication with a large number of narrow directional beams. 
\begin{algorithm}[!h] 
	\renewcommand{\thealgorithm}{Algorithm 1}
	\floatname{algorithm}{}
	\caption{$\mathbf{UCB}_{SNR}$: SNR Based Beam Selection}
	\label{alg:UCBSINR}
	\begin{algorithmic}[1]
		\STATE \textbf{Input:} $\mathcal{K}, T, T_{RSP}$
  \STATE \textbf{Initialize:} $N_k\leftarrow 0$ and $S_k\leftarrow 0$ for all $k$
		\FOR{$t=T_{RSP}+1, 2 \ldots T, $}
		    \IF {$t \leq \mathcal{K}$}
		    \STATE Select beam, $I_t = t$.
		    \ELSE 
		    \STATE $\forall k \in [\mathcal{K}]:$ compute $UCB_{k}(t)$ as given in Eq.~(\ref{equ:UCB})
		    
			\STATE Select beam, $I_t = \argmax\limits_{k \in [K]} UCB_{k}(t)$
			\ENDIF
			\STATE BS-TX transmits a data frame over $I_t$ 
   \STATE MU observes instantaneous normalized SNR, $W_{t}$
		and communicate to BS-RX over the uplink.
			\STATE $N_{I_t} \leftarrow N_{I_t}+1$ and $S_{I_t}\leftarrow S_{I_t} +W_{t}$.
		\ENDFOR 
	\end{algorithmic}
\end{algorithm}
\vspace{-4mm}
\subsection{$\mathbf{UCB}_{SNR\_AG}$: Amplitude Gated Radar-Enhanced MAB}
\label{SS:UCBAG}
In this section, we augment the $\mathbf{UCB}_{SNR}$ with the proposed radar-based target detection as shown in Fig.~\ref{fig:setup_model}b and described in Algorithm 2: $\textbf{UCB}_{SNR\_AG}$. Here, a radar target is detected in a beam when the \emph{amplitude/strength} of the scattered signal in any one or more of the range bins within the beam is above a pre-set threshold determined by CFAR. The number of beams where potential targets are detected is $\tilde{\mathcal{K}}$ (dark-colored beams in the figure) where $\mathcal{K}\geq \tilde{\mathcal{K}}$. Compared to Algorithm 1, the number of available beams is updated based on radar target detection during the first time slot (line 3 of Algorithm.2). 
Compared to $\mathbf{UCB}_{SNR}$, $\textbf{UCB}_{SNR\_AG}$ potentially offers lower regret due to the following reasons: 1) Faster target detection: The identification of the presence of targets using radar is significantly faster since returns of the scattered signals from the short-range targets are nearly instantaneous with a short round-trip delay of the order of a few $ns$. For 5G, one slot is at least 4 $ms$ assuming downlink sub-frame (1 $ms$), uplink sub-frame for reward feedback (1 $ms$), downlink (1 $ms$) and uplink data processing (1 $ms$). On average, RSP time, $T_{RSP}$ for 10 radar packets is 36 $ms$, i.e., 9 slots are sufficient to find $\tilde{\mathcal{K}}$ \cite{RadarFPGA1}; 
\begin{algorithm}[!t] 
	\renewcommand{\thealgorithm}{Algorithm 2(or)3}
	\floatname{algorithm}{}
\caption{$\textbf{UCB}_{SNR\_AG}$ (or) $\mathbf{UCB}_{SNR\_DG}$}
	\label{alg:UCBDGSINR}
	\begin{algorithmic}[1]
		\STATE \textbf{Input:} $\mathcal{K}, T, T_{RSP}$
  \STATE \textbf{Initialize:} $N_k\leftarrow 0$ and $S_k\leftarrow 0$ for all $k$
\STATE Find $\tilde{\mathcal{K}}$ using target detection (\textbf{\ref{alg:amplitude_mab}}) \\ 
\hspace{55mm}{$\vartriangleright \textbf{UCB}_{SNR\_AG}$}\\ 

Find $\vardbtilde{\mathcal{K}}$ using Doppler detection  (\textbf{\ref{alg:doppler_mab}})\\ \hspace{55mm}{$\vartriangleright\mathbf{UCB}_{SNR\_DG}$}
		     \STATE Run $\mathbf{UCB}_{SNR}$ with $\tilde{\mathcal{K}}$ arms for rest of the time horizon.\\ \hspace{55mm}{$\vartriangleright\textbf{UCB}_{SNR\_AG}$}\\
		     
		     Run $\mathbf{UCB}_{SNR}$ with $\vardbtilde{\mathcal{K}}$ arms for rest of the time horizon.\\ 
       
       \hspace{55mm}{$\vartriangleright\mathbf{UCB}_{SNR\_DG}$}
	\end{algorithmic}
\end{algorithm}
2) The proposed algorithm focuses on a subset of the total beams in which a mobile target may be present which in turn reduces the exploration time. To quantify this gain, let us fix a bandit instance. The set of beams is detected by the radar, $\Tilde{\mathcal{K}}$, is a random variable depending on the distribution of scatterers. We can assume $\tilde{\mathcal{K}}$ includes the optimal arm in each realization as it has the maximum signal strength and radar is unlikely to miss the MU. Hence the optimal arm is the same in any realized set $\tilde{\mathcal{K}}$. Expected regret over the set $\tilde{\mathcal{K}}$ is  $\mathcal{O}(\sum_{k \in \tilde{\mathcal{K}}\backslash k'} \frac{\log T}{\Delta_k})$. Clearly this bound is smaller than  $\mathcal{O}(\sum_{k \in {\mathcal{K}}\backslash k'} \frac{\log T}{\Delta_k})$ obtained for the previous case. Taking expectation over the random realizations $\tilde{\mathcal{K}}$, we get expect regret of $\textbf{UCB}_{SNR\_AG}$ as $\mathbb{E}\left[\mathcal{O}(\sum_{k \in {\tilde{\mathcal{K}}}\backslash k'} \frac{\log T}{\Delta_k})\right] \leq \mathcal{O}(\sum_{k \in {\mathcal{K}}\backslash k'} \frac{\log T}{\Delta_k})$. Thus $\textbf{UCB}_{SNR\_AG}$ is better than that of $\mathbf{UCB}_{SNR}$ resulting in an improvement in the performance. 
\vspace{-3mm}
\subsection{$\mathbf{UCB}_{SNR\_DG}$: Doppler Gated Radar-Enhanced MAB}
In the $\textbf{UCB}_{SNR\_AG}$ algorithm, all the beams where radar targets are present are selected. However, some of the beams correspond to SCS as shown in Fig.\ref{fig:setup_model}c. The SCS is distinguished to be of two types: some give rise to direct scattering at the radar (termed $SCS_1$)  with zero Doppler while others give rise to Doppler-shifted returns at the radar through multipath with respect to the MU (termed $SCS_2$). 
The proposed Doppler-enhanced MAB algorithm is described in Algorithm 3: $\mathbf{UCB}_{SNR\_DG}$. Since direct path returns from SCS do not give any type of information regarding the MU, they can be excluded from the list of candidate beams based on the Doppler shift estimated from the radar signal processing described earlier. 
\begin{algorithm}[!t] 
	\renewcommand{\thealgorithm}{Subroutine 1}
	\floatname{algorithm}{}
	\caption{$\mathbf{AG}:$  $\tilde{\mathcal{K}}$ Beams Selection Using Radar Target Detection}
	\label{alg:amplitude_mab}
	\begin{algorithmic}[1]
		\STATE \textbf{Input:} $\mathcal{K}$
		\STATE \textbf{Output:} $\tilde{\mathcal{K}}$ 
	
		\FOR{$\theta =1, 2 \ldots \mathcal{K}, $}
		   	     
		    \STATE $\mathbf{\chi}_{\theta}$: Matched filtering across fast time samples 
		    
			\STATE CFAR detection:
               \IF {$\mathbf{\chi}_{\theta} \geq \gamma$}
		    \STATE Include beam, $\theta$ in subset $\tilde{\mathcal{K}}$
		    \ENDIF
			
		\ENDFOR
	\end{algorithmic}
\end{algorithm}
Hence the total number of beams where potential MU are detected is $\vardbtilde{\mathcal{K}}$ and $\mathcal{K}\geq\tilde{\mathcal{K}}\geq \vardbtilde{\mathcal{\mathcal{K}}}$. Thus, fewer candidate beams (dark-colored beams in the figure) result in lower exploration time. The theoretical explanation for the reduction in beams follows the same logic provided for the previous algorithm and hence is not repeated here. 
Note that static communication targets are not likely to have first triggered the necessity for the selection of a new beam by the BS and hence, static targets can be interpreted safely as SCS.
\begin{algorithm}[!ht] 
	\renewcommand{\thealgorithm}{Subroutine 2}
	\floatname{algorithm}{}
	\caption{ $\mathbf{DG}$: $\kd$ Beams Selection Based on Doppler Estimation }
	\label{alg:doppler_mab}
	\begin{algorithmic}[1]
		\STATE \textbf{Input:} $\mathcal{K}$ 
		\STATE \textbf{Output:} $\vardbtilde{K}$ 
	
		\FOR{$\theta =1, 2 \ldots \mathcal{K}, $}		    
			\STATE 1D-MUSIC for $\hat{r}_a$ across Q packets 
               \IF {$\hat{f}_a \neq 0$} 
		    \STATE Include beam, $\theta$, in subset $\vardbtilde{\mathcal{K}}$ 
		    \ENDIF			
		\ENDFOR
	\end{algorithmic}
\end{algorithm}
 \vspace{-2mm}
\section{Performance Analysis}
\label{sec:Simulation exp} 
We consider a three-dimensional (3D) Cartesian coordinate space with the ground plane defined by the $x$ and $y$ axes and the height axis along $z$. The BS is located at $[0,0,0]$ m with a $y$-aligned uniform linear array (ULA) of 32 antennas with an antenna spacing of $\lambda/2$ where $\lambda$ is the wavelength corresponding to the center frequency $f_c$ of 60 GHz. We adopt a 16QAM modulation and coding scheme with 512 OFDM subcarriers with a signal bandwidth of 1.76 GHz.
We assume that the channel consists of a single MU and multiple SCS as shown in Fig.~\ref{fig:setup_model}. The radar scattered returns of each of these are confined to a single beam. We assume that the MU is also with a 32-element ULA and is initially located at $[50,20,0]$m and subsequently moves with a constant velocity of $v$ m/s along the $x$ axis. Both MU and SCS are modeled as isotropic point scatterers and the SCS are distributed randomly across the 3D Cartesian space. 

\begin{figure}[htbp]
    \centering    \includegraphics[scale=0.3]{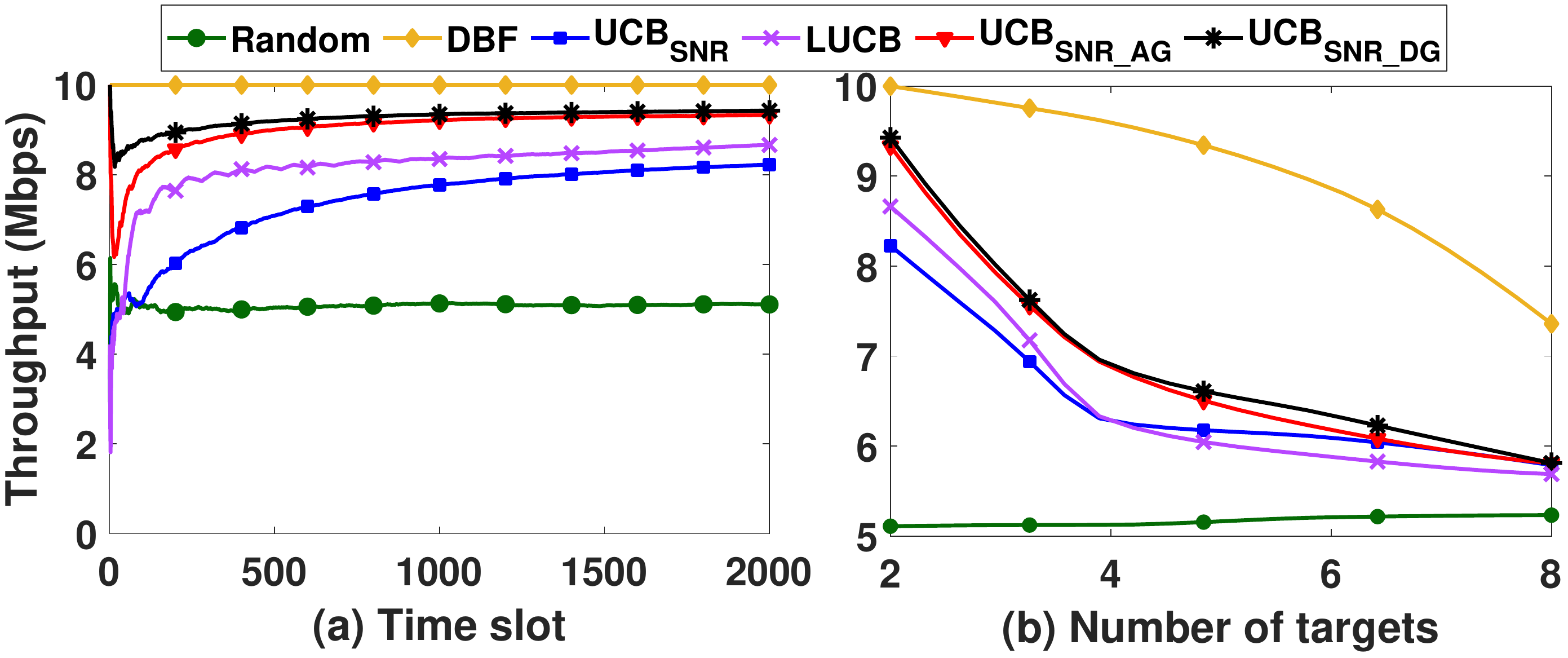}
    \vspace{-2mm}
    \caption{\footnotesize (a) Throughput at different time slots for two targets, and (b) Throughput at the end of the horizon for different numbers of targets.}
    \label{fig:target analysis}
   \vspace{-1mm}
\end{figure}

\begin{figure}[htbp]
    \centering
    \includegraphics[scale=0.3]{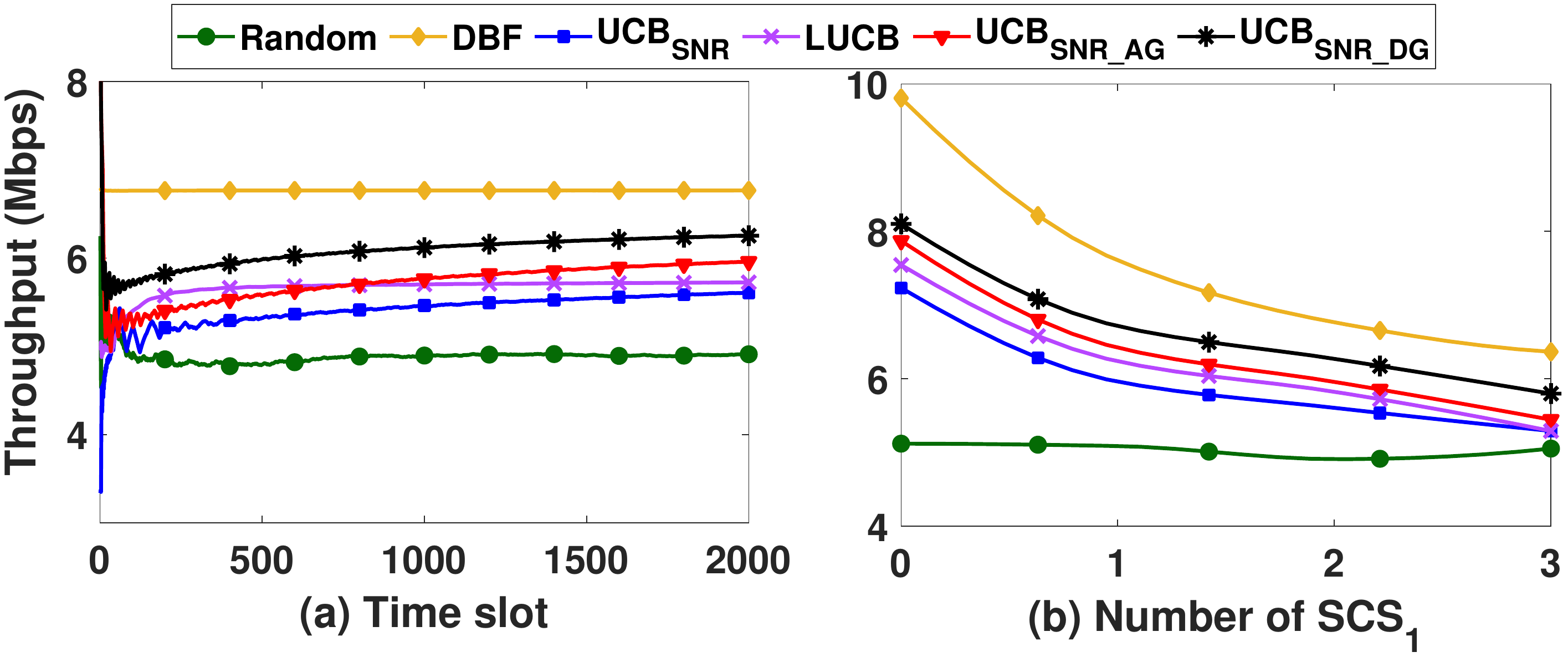}
   \vspace{-2mm}
    \caption{\footnotesize (a) Throughput at different time slots for four targets (including two $SCS_1 $, one $SCS_2$ and one MU, and (b) Throughput at the end of the horizon for different number of $SCS_1$.}
    \label{fig:blocker analysis}
    \vspace{-2mm}
\end{figure}

The throughput, $\Upsilon$, is calculated as $\left(1 - \frac{\sum_{i=1}^{N_t}BER_{i}}{N_t}\right)\frac{D}{T_d}$ where, $BER_i$ corresponds to the bit error rate of $i^{th}$ time slot, $D$ and $T_d$ correspond to the total number of bits and the time duration for each slot respectively.
We benchmark the $\Upsilon$ performance of the proposed algorithms, $\textbf{UCB}_{SNR\_AG}$ and $\mathbf{UCB}_{SNR\_DG}$ with conventional  $\mathbf{UCB}_{SNR}$, lower upper confidence bound (LUCB) described in \cite{kalyanakrishnan2012pac}, digital beamforming and trivial random beam selection approach.  We present the effects of the number of targets, number of beams, the Doppler velocity resolution, and radar receiver SNR on $\Upsilon$. Each result presented in this section is obtained after averaging over 15 independent experiments and each experiment's duration/horizon is 2000 time slots. \\
\noindent \textbf{Effect of Number of Radar Scatterers:} In Fig.~\ref{fig:target analysis}(a), we compare $\Upsilon$ of all the algorithms at different instants of the time horizon. Here, we assume one MU and vary the number of SCS. The Doppler velocity of the MU is fixed to 3 m/s and the angular resolution is $4^o$ resulting in a total of 41 candidate beams spanning from $-80^o$ to $80^o$ and the Doppler velocity resolution is 1 m/s. 
It can be observed in the figure that the proposed algorithms offer higher $\Upsilon$ than the benchmarked approaches - except for DBF - due to faster identification of the optimal beam. DBF provides the best-case results since all the beams are tested simultaneously. However, this approach is not pursued since the implementation of multiple synchronized receiver chains is costly and complex. In Fig.~\ref{fig:target analysis}(b), we compare  $\Upsilon$ of all algorithms at the end of the horizon  as the number of SCS increases. We observe that the performance of all MAB-based approaches is significantly better than the random selection approach validating the need for a learning algorithm. As expected, the difference between $\mathbf{UCB}_{SNR}$ (non-radar based beam selection) and $\textbf{UCB}_{SNR\_AG}$ or $\mathbf{UCB}_{SNR\_DG}$ (radar-based beam selection) reduces as the number of targets increases. This is because when there are a large number of SCS, there is a proportionate increase in the number of candidate beams. Further, we can observe that there is a slight improvement in $\Upsilon$ for $\mathbf{UCB}_{SNR\_DG}$ than $\textbf{UCB}_{SNR\_AG}$ as beams with zero Doppler get eliminated.\\
\noindent \textbf{Effect of Angular Resolution} Next, we fix the number of targets to three, which includes one MU and two SCS, and compare the performance for different angular resolutions or the number of candidate beams. The rest of the parameters are the same as Fig.~\ref{fig:target analysis}. 
\begin{figure}[htbp]
    \centering
    \includegraphics[scale=0.3]{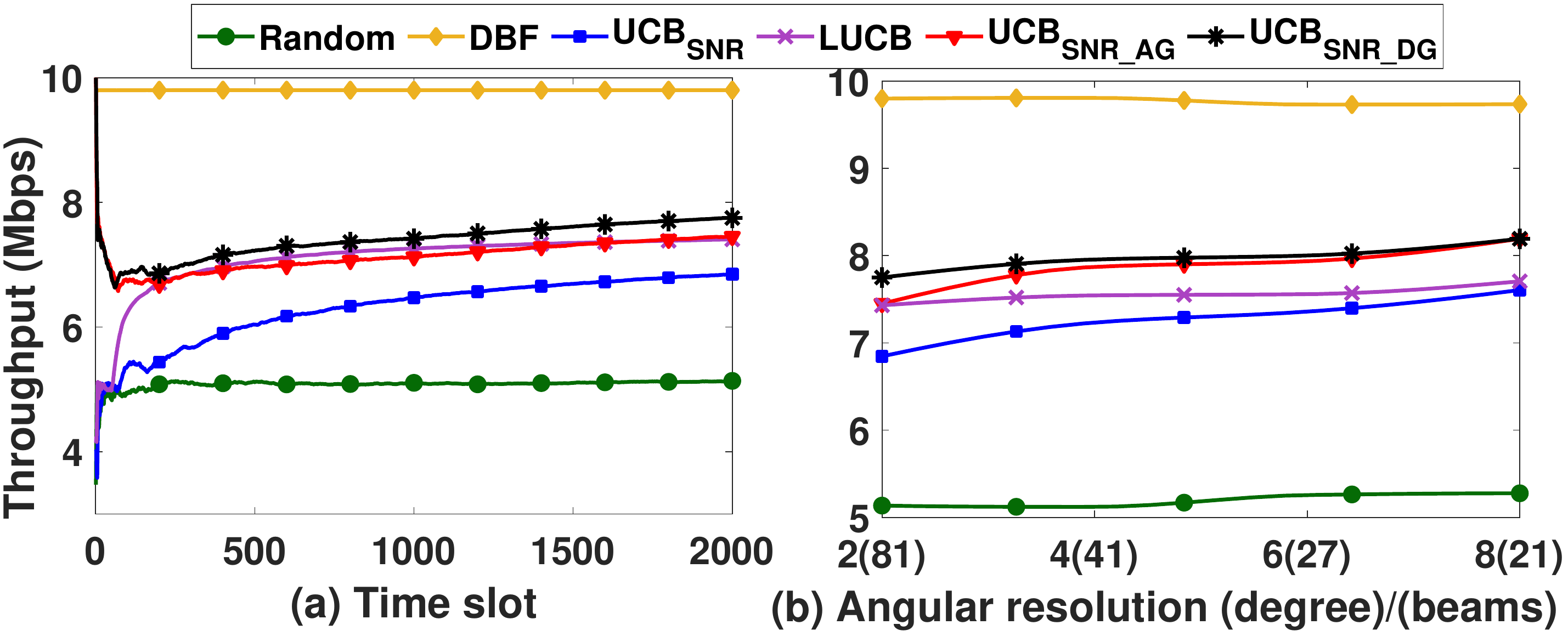}
    \vspace{-2mm}
    \caption{\footnotesize (a) Throughput at different time slots for angular resolution of $2^o$, and (b) Throughput at the end of the horizon for different angular resolutions/number of candidate beams.}
    \label{fig: Angular resolution analysis}
    \vspace{-2mm}
\end{figure}
In  Fig.~\ref{fig: Angular resolution analysis}(a), we compare the $\Upsilon$ for angular resolution of $2^o$ and it can be observed that the proposed approaches offer higher  $\Upsilon$ than $\mathbf{UCB}_{SNR}$ and random approaches at all time slots. In Fig.~\ref{fig: Angular resolution analysis}(b), we compare the $\Upsilon$ at the end of the horizon for different numbers of candidate beams. As expected,  $\Upsilon$ increases as the number of beams reduces due to lower exploration time. However, proposed  $\textbf{UCB}_{SNR\_AG}$ or $\mathbf{UCB}_{SNR\_DG}$ algorithms significantly outperform $\mathbf{UCB}_{SNR}$ and random approaches in all cases. The $\mathbf{UCB}_{SNR\_DG}$ offers higher  $\Upsilon$ than $\textbf{UCB}_{SNR\_AG}$ for smaller angular resolution (i.e., large $\mathcal{K}$) due to more opportunities of beam elimination. However, the impact of Doppler processing reduces with the increase in angular resolution due to the occurrence of fewer beams.\\
\noindent \textbf{Effect of Velocity Resolution:} Next, we study the effect of Doppler velocity resolution on $\Upsilon$. The simulation is carried out for Doppler velocity resolution of 1 m/s, 2 m/s, 3 m/s, 4 m/s and 5 m/s. We assume the MU with a Doppler velocity of 3 m/s and three targets (one MU and two SCS). 
\begin{figure}[htbp]
    \centering
    \includegraphics[scale=0.31]{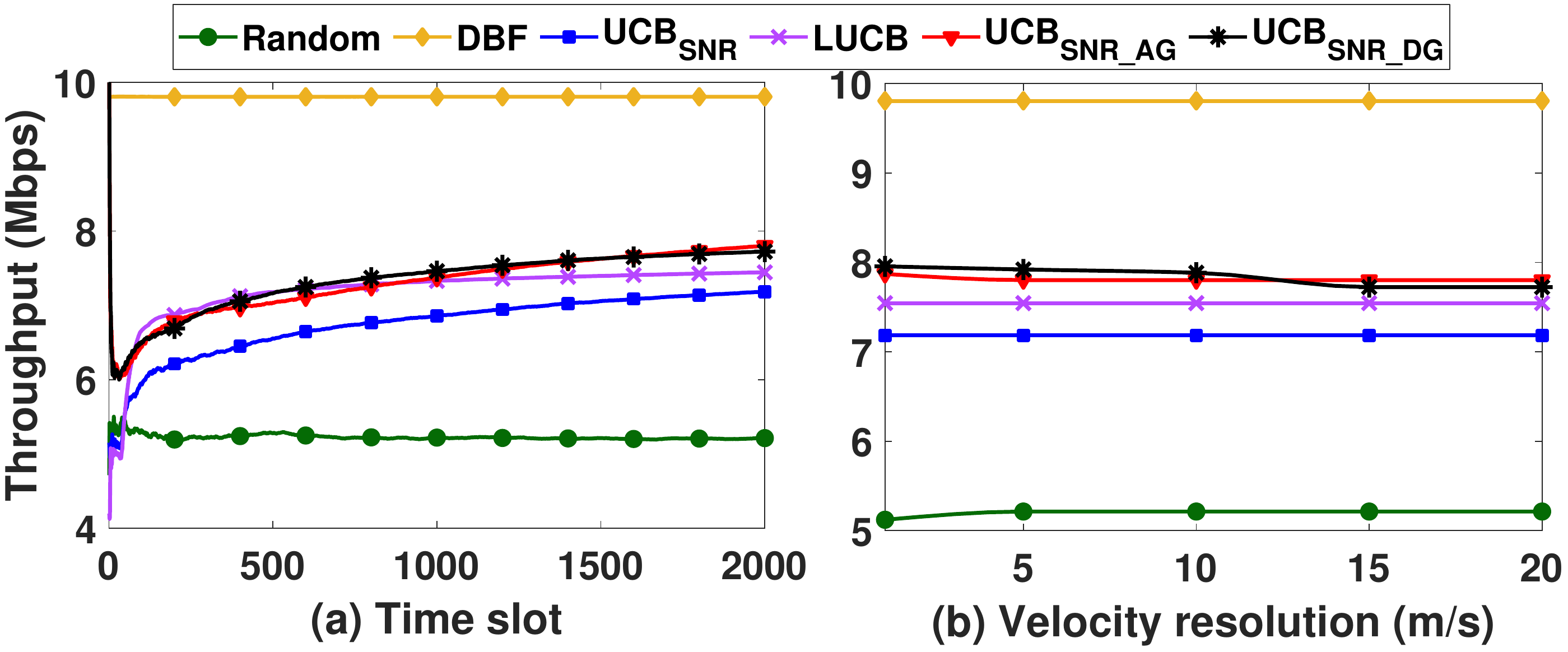}
   \vspace{-2mm}
    \caption{\footnotesize (a) Throughput at different time slots for velocity resolution of $15$m/s, and (b) Throughput at the end of the horizon for different velocity resolutions.}
    \label{fig:Doppler resolution analysis}
\end{figure}
In Fig.~\ref{fig:Doppler resolution analysis}(a), we consider the velocity resolution of 5 m/s, which is not sufficient to detect the MU and hence, the performance of $\mathbf{UCB}_{SNR\_DG}$ is poorer than that of $\textbf{UCB}_{SNR\_AG}$. As shown in Fig.~\ref{fig:Doppler resolution analysis}(b), the velocity resolution of radar signal processing should be carefully chosen. Thus, $\mathbf{UCB}_{SNR\_DG}$ may not offer better performance than $\textbf{UCB}_{SNR\_AG}$ even though it incurs higher computational cost due to Doppler estimation. As expected,  $\Upsilon$ of $\mathbf{UCB}_{SNR}$ and random approaches are independent of the velocity resolution since they do not use this information for beam selection. \\ 
\noindent \textbf{Effect of Radar Receiver SNR:} Further, we analyze all the algorithms for four different SNRs at the radar receiver ranging from -10 dB to 10 dB. Here, also we simulate three targets (one MU and two SCS), and the velocity resolution is set as 1 m/s. 
\begin{figure}[htbp]
    \centering
    \includegraphics[scale=0.3]{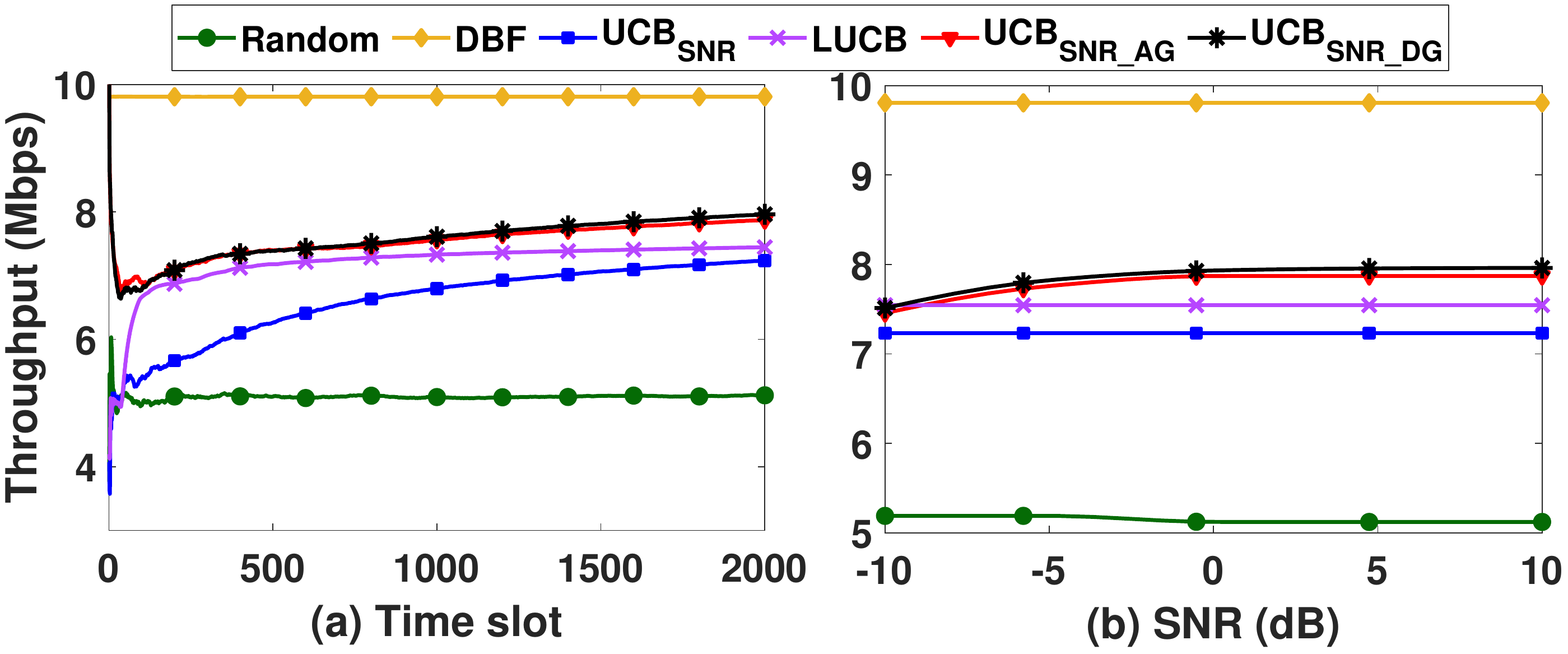}
    \vspace{-2mm}
    \caption{\footnotesize (a) Throughput at different time slots for 10dB SNR, and (b) Throughput at the end of the horizon for different SNRs.}
    \label{fig:SNR analysis}
   \vspace{-1mm}
\end{figure}
Fig.~\ref{fig:SNR analysis}(a) represents the viewgraph of  $\Upsilon$ with respect to time slots for a high SNR (10 dB). We observe in Fig.~\ref{fig:SNR analysis}(b) that the regret improves for the $\textbf{UCB}_{SNR\_AG}$ and $\mathbf{UCB}_{SNR\_DG}$ with increase in SNR. At lower SNR, the poor prediction of the target presence amidst noise results in higher regret. Note that since we consider the SNR with respect to the radar receiver, change in SNR does not have any impact on the random beam selection as well as $\mathbf{UCB}_{SNR}$ algorithms.\\
\noindent \textbf{Impact of Doppler Processing:}
Further, we discuss the scenario where the number of $SCS_1$ is greater than $SCS_2$ such that the candidate beams selected by the $\mathbf{UCB}_{SNR\_DG}$ are significantly lesser than those selected by the $\textbf{UCB}_{SNR\_AG}$. In Fig.~\ref{fig:blocker analysis}(a), we consider one MU, one $SCS_2$ and two $SCS_1$.  The angular resolution is $4^o$ and the velocity resolution is 1 m/s. 
Since $SCS_2$ gives rise to Doppler shift at the BS through multipath from the MU, the corresponding beam is retained for further scanning while the beams corresponding to the two $SCS_1$ that give rise to zero-Doppler shift are excluded in $\mathbf{UCB}_{SNR\_DG}$. This results in better performance of the algorithm compared to $\mathbf{UCB}_{SNR\_AG}$. Further, the improvement is greater as the number of $SCS_1$ increases for a fixed number of $SCS_2$ as seen in Fig.~\ref{fig:blocker analysis}(b).
\section{Conclusion}
\label{sec:Conclusion}
In this work, we demonstrate how radar-enhanced MAB within a JRC BS can substantially reduce the exploration time by selecting only those candidate beams that detect the presence of radar targets of which the MU may be one. Further reduction in the exploration time is realized by distinguishing SCS from MU through radar-based Doppler estimation. Simulation results demonstrate an overall improvement in the communication link metrics with the reduction in the exploration time  through radar-enhanced MAB when compared with conventional MAB algorithms. 

\bibliographystyle{ieeetran}
\bibliography{main}
\end{document}